\documentclass[apj]{emulateapj}
\usepackage{natbib}
\usepackage{mathtools}
\usepackage{float}

\shorttitle{The age of M\,15}
\shortauthors{Monelli et al.}

\begin{document}

\title{The absolute age of the globular cluster M\,15 using near-infrared 
adaptive optics images from PISCES/LBT. }

\author{
M. Monelli\altaffilmark{1,2},
V. Testa\altaffilmark{3},
G. Bono\altaffilmark{4,3},
I. Ferraro\altaffilmark{3},
G. Iannicola\altaffilmark{3},
G. Fiorentino\altaffilmark{5},
C. Arcidiacono\altaffilmark{5,6},
D. Massari\altaffilmark{5,8},
K. Boutsia\altaffilmark{3},
R. Briguglio\altaffilmark{6},
L. Busoni\altaffilmark{6},
R. Carini\altaffilmark{3},
L. Close\altaffilmark{9},
G. Cresci\altaffilmark{5},
S. Esposito\altaffilmark{6},
L. Fini\altaffilmark{6},
M. Fumana\altaffilmark{10},
J.C. Guerra\altaffilmark{11},
J. Hill\altaffilmark{11},
C. Kulesa\altaffilmark{9},
F. Mannucci\altaffilmark{6},
D. McCarthy\altaffilmark{9},
E. Pinna\altaffilmark{6},
A. Puglisi\altaffilmark{6},
F. Quiros-Pacheco\altaffilmark{6},
R. Ragazzoni\altaffilmark{12},
A. Riccardi\altaffilmark{6},
A. Skemer\altaffilmark{9},
M. Xompero\altaffilmark{6},
}

\altaffiltext{*}{Observations were carried out using the Large Binocular
Telescope at Mt. Graham, AZ. The LBT is an international collaboration among
institutions in the United States, Italy, and Germany. LBT Corporation partners are
The University of Arizona on behalf of the Arizona university system; Istituto
Nazionale di Astrofisica, Italy; LBT Beteiligungsgesellschaft, Germany, representing
the Max-Planck Society, the Astrophysical Institute Potsdam, and Heidelberg
University; The Ohio State University; and The Research Corporation, on behalf of
The University of Notre Dame, University of Minnesota, and University of Virginia.}

\altaffiltext{1}{Instituto de Astrof\'{i}sica de Canarias, La Laguna, Tenerife, Spain; monelli@iac.es}
\altaffiltext{2}{Departamento de Astrof\'{i}sica, Universidad de La Laguna, Tenerife, Spain}
\altaffiltext{3}{INAF-Osservatorio Astronomico di Roma, Via Frascati 33, 00040 Monteporzio Catone, Italy}
\altaffiltext{4}{Universit{\'a} di Roma "Tor Vergata", Via della Ricerca Scientifica 1, 00133 Roma}
\altaffiltext{5}{INAF-Osservatorio Astronomico di Bologna, Via Ranzani 1, I-40127 Bologna, Italy}
\altaffiltext{6}{INAF-Osservatorio Astronomico di Arcetri, Largo Enrico Fermi 5, I-50125 Firenze, Italy}
\altaffiltext{7}{Dipartimento di Fisica e Astronomia, Università degli Studi di Bologna, v.le Berti Pichat 6/2, I-40127 Bologna, Italy}
\altaffiltext{8}{Kapteyn Astronomical Institute, Landleven 12, NL-9747 AD Groningen, the Netherlands}
\altaffiltext{9}{Steward Observatory, Department of Astronomy, University of Arizona, Tucson, AZ 85721, USA}
\altaffiltext{10}{INAF-IASF, via E. Bassini 15, 20133 Milano, Italy}
\altaffiltext{11}{LBT Observatory, Univ. of Arizona, 933 North Cherry Ave., Tucson, AZ 85721, USA}
\altaffiltext{12}{INAF-Oservatorio Astronomico di Padova, Vicolo dell'Osservatorio 5, 35122, Padova, Italy}

%
\begin{abstract}

We present deep near-infrared (NIR) $J$, $K_{\mathrm{s}}$ photometry of the old,
metal-poor Galactic globular cluster M\,15 obtained with images collected with
the LUCI1 and PISCES cameras available at the Large Binocular Telescope (LBT).
We show how the use of First Light Adaptive Optics system coupled with the
(FLAO) PISCES camera allows us to improve the limiting magnitude by $\sim$2 mag in
$K_{\mathrm{s}}$. By analyzing archival HST data, we demonstrate that the
quality of the LBT/PISCES color magnitude diagram is fully comparable with
analogous space-based data. The smaller field of view is balanced by the
shorter exposure time required to reach a similar photometric limit. \\ We
investigated the absolute age of M\,15 by means of two methods: {\it i)} by
determining the age from the position of the main sequence turn-off; and {\it ii)}
by the magnitude difference between the MSTO and the well-defined knee detected
along the faint portion of the MS. We derive consistent values of the absolute
age of M\,15, that is 12.9$\pm$2.6 Gyr and 13.3$\pm$1.1 Gyr, respectively.

\end{abstract}



\section{Introduction}\label{sec:intro}


Cosmological results based on recent Cosmic microwave background (CMB) experiments
(Boomerang, WMAP, PLANCK), on Baryonic Acoustic oscillations
\citep[BAO][]{eisenstein05}, on supernovae observations \citep{riess98,riess11} 
and on gravitational lensing \citep[]{suyu10,suyu13} opened the path to the era of
precision cosmology. However, the quoted experiments are affected by  an intrinsic
degeneracy in the estimate of cosmological parameters, e.g. the Hubble constant
H$_0$.  To overcome this problem either specific priors or the results of
different experiments are used \citep{bennett14}.

Recent evaluations of the H$_0$ based on CMB provide values ranging 
from 70.0$\pm$2.2 km s$^{-1}$ Mpc$^{-1}$ \citep[WMAP9][]{hinshaw13} to  
67.8$\pm$0.9 km s$^{-1}$ Mpc$^{-1}$  \citep{planck15}. Similar values have also 
been obtained by BAO plus supernovae using the so-called inverse distance
ladder  suggesting a value of 68.6$\pm$2.2 km s$^{-1}$ Mpc$^{-1}$ 
\citep{cuesta15}. On the other hand, resolved objects (Cepheids plus
supernovae) provide  H$_0$ values ranging from 73$\pm$2 (random) $\pm$4
(systematic)  km s$^{-1}$ Mpc$^{-1}$ \citep{freedman10} to  73.8$\pm$2.4 km
s$^{-1}$ Mpc$^{-1}$ \citep{riess11}. Slightly larger  values of the Hubble
constant were obtained by \citet{suyu13} using  gravitational lens time delays
(80.0$^{+4.5}_{-4.7}$  km s$^{-1}$ Mpc$^{-1}$,  uniform $H_0$ in flat
$\Lambda$CDM). 

The above estimates of the Hubble constant indicate that there is some tension
between the results based on CMB and BAO and those based on primary and
secondary distance indicators. This critical issue has been addressed in 
several recent papers suggesting a difference that range from almost 2$\sigma$
\citep{efstathiou14} to more than 2.5$\sigma$ \citep{riess11}. The quoted 
uncertainties on the Hubble constant open the path to new physics concerning 
the number of relativistic species and/or the mass of neutrinos
\citep{dvorkin14,wyman14}. Moreover and even more importantly, the above range
in $H_0$ implies an  uncertainty on the age of the universe --$t_0$-- of the
order of 2 Gyr. This uncertainty has a substantial impact not only on galaxy
formation and evolution, but also on the age of the most ancient stellar
systems, i.e. the globular clusters (GCs).  

The absolute age of GCs can be independently estimated using stellar
astrophysics and it is affected by theoretical, empirical 
and intrinsic uncertainties. 

{\em Theoretical}-- Stellar evolutionary models adopted to construct cluster 
isochrones are affected by uncertainties in the input physics. In particular, in
the adopted micro (opacity, equation of state, astrophysical screening  factors)
and in macro-physics (mixing length, mass loss, atomic diffusion  radiative
levitation, color-temperature transformations). The impact that  the quoted
ingredients have on cluster isochrones have been discussed in  detail in the
literature \citep{vandenberg13, pietrinferni04, pietrinferni09,cassisi14}. The typical 
uncertainty in the adopted clock --the main sequence Turn Off (MSTO)--  is of
the order of 10\%. Thus suggesting that theoretical uncertainties  does not
appear to be the dominant source in the error budget of the absolute age of
GCs. 

{\em Empirical}-- The main source of uncertainty in the absolute age estimate 
of globular clusters are the individual distances ($\Delta \mu_0 \sim$0.1 mag 
in the true distance modulus implies an uncertainty of 1 Gyr in 
the absolute age). The age estimate is even more affected when the uncertainties 
in the reddening correction and in the reddening law are taken into account
\citep{stetson14a}. 

Importantly, the massive use of multi-object fiber spectrographs provided the
opportunity to construct a firm metallicity scale including a significant
fraction of GGCs \citep{carretta09}, thus reducing the uncertainties in the
iron and in the $\alpha$-element abundances.   

{\em Intrinsic}-- Dating back to more than forty years ago, spectroscopic 
investigations brought forward a significant star-to-star variation in 
C and in N among cluster stars \citep{osborn71}. This evidence was soundly 
complemented by variation in Na, Al, and in O \citep{cohen78, 
pilachowski83, leep86} an by 
anti-correlations in CN--CH \citep{kraft94} and  
in O--Na and Mg--Al \citep{suntzeff91,gratton12}. 

The above evidence has further strengthened by the occurrence of multiple 
stellar populations in more massive clusters \citep{bedin04,piotto05,piotto07}.
However, detailed investigations  concerning the different stellar populations
indicate a difference in age  that is, in canonical GCs, on average shorter than
1~Gyr  \citep{ventura01,cassisi08}. The intrinsic uncertainty does not seem to
be the main source of the error budget of the GCs absolute age.

To overcome or to alleviate the quoted uncertainties, have been suggested 
different approaches mainly based on relative age estimates, the so-called 
vertical and horizontal methods \citep{marinfranch09,dotter11,vandenberg13}.
In this context the relative age is estimated as a difference between the clock
(the MSTO) and an evolved reference point either the horizontal  branch or a
specific point along the red giant branch. The key advantage of these  methods
is that they are independent of uncertainties on cluster distance and 
reddening. However, they rely on the assumption that the reference points are 
independent of cluster age and introduce new theoretical uncertainties 
(conductive opacities, \citealt{cassisi07}; extra-deep mixing along the RGB, 
\citealt{denissenkov04}). It goes without saying that the transformation of 
relative ages in absolute ages using a calibrating GC introduces the typical 
uncertainties already discussed. 

More recently it has been suggested to use  as a reference point  a well defined
knee along the low-mass regime  of the main sequence (MSK). The MSK has already
been detected in several old  ($\omega$~Cen, \citealt{pulone98}; M4,
\citealt{pulone99,milone14b,braga15};  NGC~3201, \citealt{bono10a}; 47~Tuc,
\citealt{lagioia14}; NGC~2808,  \citealt{milone12a}; M71,
\citealt{dicecco15}) and intermediate-age \citep{sarajedini09b,an09a,an09b}
stellar  systems and in the Galactic bulge \citep{zoccali00a} by using either
near-infrared (NIR) and/or optical NIR CMDs.

Nevertheless, one of the most difficult
observational problem in measuring stellar magnitudes and colors in GCs is that
they are intrinsically crowded stellar systems, and therefore the photometry of
their stars is strongly limited by poor weather conditions. In particular, bad
seeing (larger than $\sim$1$\arcsec$) has the effect of severely limiting the 
identification and measurement of faint stars. This means a systematic increase 
in the limiting magnitudes and in the photometric accuracy when moving from the 
outskirts to the innermost cluster center.

Twenty five years ago the advent of the Hubble Space Telescope (HST) started a new
era, and the high spatial resolution provided by space images collected in
optical bands allowed us to resolve the core of GCs. Recently, a similar
resolution is becoming possible from ground based  observations using NIR
cameras available on 10m class telescopes assisted by  adaptive optics (AO)
systems. This technology allows ground-based observations reach the
diffraction limit over a modest field of view  ($\sim$ 1$\times$1$\arcmin$).
High-resolution NIR images of GCs can have a relevant impact on current 
astrophysical problems as soundly demonstrated by MAD  (Multi-Conjugated
Adaptive Optics Demonstrator, \citealt{marchetti03}) the pilot VLT instrument
built to test on the sky the feasibility of a multi-conjugated AO
(MCAO) systems \citep{ferraro09,bono09,moretti09,fiorentino11}. The robustness
of the current  MCAO systems has been further supported by GeMS/GSAOI available
at the  GEMINI-South telescope \citep{neichel14a,neichel14b,rigaut14}. This new 
system uses both natural guide stars for the tip tilt correction and  five
artificial stars to close the loop and it has been able to delivery  uniform NIR
images approaching its diffraction limit in Galactic bulge \citep{saracino15}
and halo (Turri et al. 2015) GCs.

\begin{centering}
\begin{figure}
 \includegraphics[width=8cm]{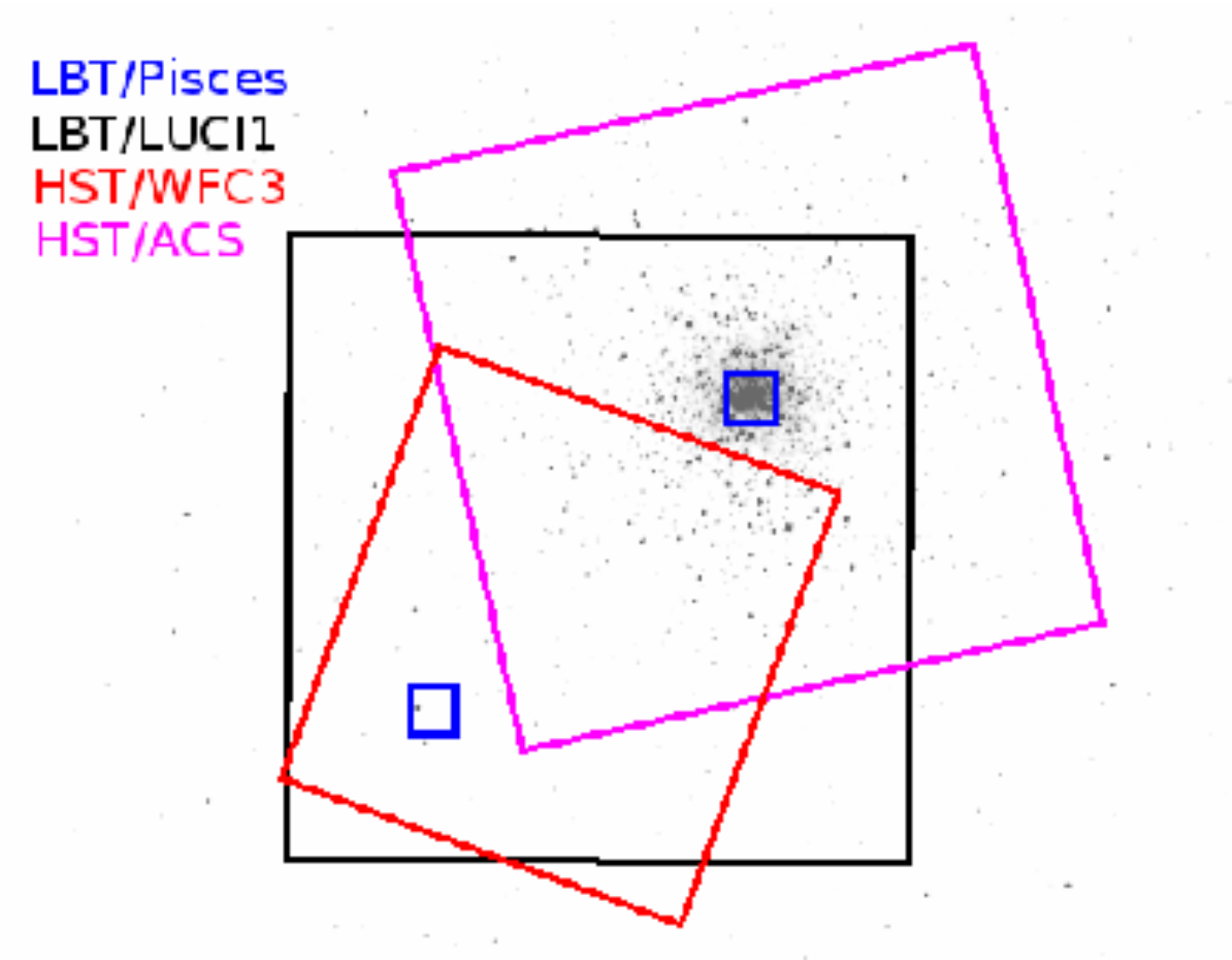}
 \caption{SDSS image of M\,15 with superimposed the 
 fields of the different data sets collected for this project. Detailed
 analisys of the central PISCES field will be presented in a forthcoming paper.}
 \label{fig:map}
\end{figure}
\end{centering}

In this context we have collected AO images of the GC M\,15 (NGC\,7078), using the
First Light Adaptive Optics \citep[FLAO,][]{esposito10} system mounted on the
Large Binocular Telescope (LBT). This cluster is located at $\approx$10 Kpc (10.4,
\citealt{durrell93}; 9.9 kpc, \citealt{mcnamara04}; 10.4 kpc,
\citealt{vandenbosch06}) and is affected by moderate interstellar extinction
(E($B-V$)=0.08 mag, \citealt{sandage81}; 0.10, \citealt{schlafly11}, 0.12,
\citealt{schlegel98}). Most interestingly, it is among the most metal-poor
Galactic GCs ([Fe/H] $\approx -2.4$, \citealt{kraft03}), and therefore it possibly
traces the oldest component of our Galaxy (see Table 1 for the parameters
assumed  in our analysis). Notably, despite multiple populations have been proved
to exist in this cluster \citep{monelli13,piotto15}, so far there is no evidence
of multiple turn-off or sub-giant branches as for NGC\,1851 \citep{milone08}
that could affect the age estimate.

\section{Optical and NIR data sets}\label{sec:data}

{The present work uses four different data sets from different imagers. In the
following we summarize the main properties of each of them. A summary
is given in Table 2}.

	\subsection{LBT/PISCES data}

PISCES is an near-IR imager covering a wavelength range 1-2.5 $\mu$m with an
Hawaii 1024 px$\times$1024 px HgCdTe array, installed at the front
bent-Gregorian focus \citep{mccarthy01} of the LBT. It critically samples a
diffraction-limited PSF with a plate scale of 0.0193$^{\arcsec}$/px.
Observations were carried using PISCES together with the FLAO system mounted on
the DX (right) telescope of LBT. M\,15 was observed on October 14-15 2011 during
the Science Verification Time for the FLAO system.  This is a twin of
672-actuators voice-coil based, contactless adaptive mirror
\citep{salinari94,davies10} controlled by the means of a pyramid 
\citep{ragazzoni96} wavefront sensor. The FLAO uses solely Natural Guide Star as
reference and it retrieves high Strehl-ratio over a broad wavelength range,
reaching peak performance on bright reference in the NIR (80\% in the H band).
Once at regime phase the two FLAO systems will feed the LUCI1 and LUCI2 cameras
\citep{lefevre03}.

 \begin{table}[ht!]
 \begin{center}
 \caption{Basic parameters of M\,15.}
 \begin{tabular}{ccc}
 \hline
 \hline
  (m-M)$_0$ [mag]  & 15.14  & \citet{harris96,durrell93}  \\
  E(B-V) [mag]     &  0.08  & \citet{sandage81}  \\
  $[$Fe/H$]$       & -2.4   & \citet{kraft03}  \\
 \hline
 \hline
 \end{tabular}
 \end{center} 
 \label{tab:tab01}
 \end{table}


Two fields were acquired with the FLAO+PISCES setup, one centered on the cluster
core, the other approximately 3 arcmin South-West of the cluster center. In both
cases the selection criterion for the field was the presence of a suitable star
for wavefront analysis in the field-of-view, with magnitude $R = 12.6$ mag and$R
= 12.9$ mag for the central and outer field, respectively. In the current
investigation we will focus on the external field.  A preliminary photometric
analysis of the central field together with  a detailed investigation of the
variation of the PSF across the field  of view has already been discussed by
\citet{fiorentino14b}. A comprehensive    analysis will be addressed in a
forthcoming investigation. During the observations, weather conditions were
photometric with good natural seeing conditions
(0.65$^{\arcsec}$--0.9$^{\arcsec}$, as recorded by the DIMM). The AO allowed to
reach a mean FWHM of 0.05$^{\arcsec}$ and of 0.06$^{\arcsec}$ in the J-- and in
the  $K_{\mathrm{s}}$--band, respectively, as measured on the images. The Strehl
ratio on the  quoted images reached 28\% (J) and 60\% ($K_{\mathrm{s}}$)
consistently with the expected scaling vs. wavelength.

	\subsection{LBT/LUCI1 data}

One pointing with the spectro-imager LUCI1 at LBT was collected for 
calibration purposes.  The set of observations was secured on June, 
21-22 2012 in the $J$ and $K_{\mathrm{s}}$ filters, under good seeing 
conditions ($\sim$0.7$^{\arcsec}$). The data set was taken with the 
center of the cluster in the NE quadrant of the image in order to 
include both the central and the outer field observed with PISCES.
We adopted this observing strategy to constrain possible systematics 
in the absolute calibrations due to positional effects.
An off-source set of images was also taken to perform median-sky subtraction
and superflat construction (see Section \ref{sec:reduction}).

	\subsection{HST optical and NIR data}

Complementary data sets will be used in the analysis of the LBT images. In
particular, a series of nine images has been retrieved from the HST archive. 
They were collected with the WFC3 in the $F160W$ passband.  Furthermore, we 
will make use of  $F606W$ and $F814W$ photometry of M\,15,  retrieved from the
``ACS Survey of Galactic Globular Clusters'' database \citep{sarajedini07}.

\begin{centering}
\begin{figure*}
 \includegraphics[width=17cm]{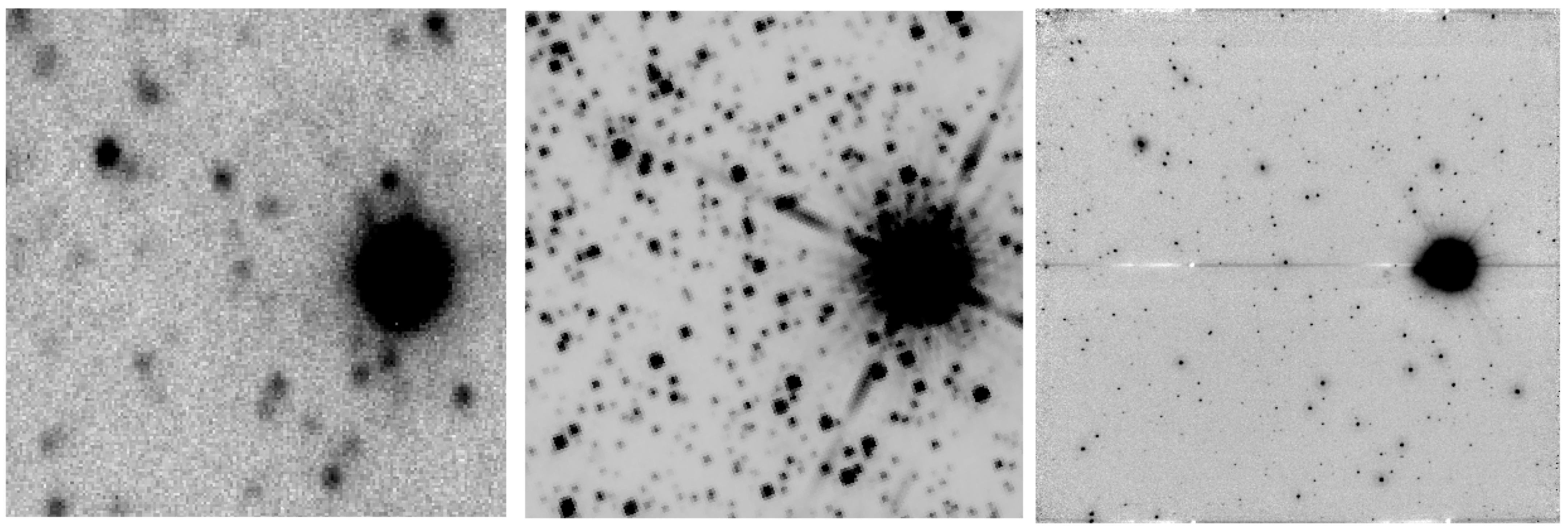}
 \caption{The sky region covered by the PISCES camera (right panel) as seen
 also by LUCI1 (left) and the WFC3 (center). Note the change in spatial resolution,
 from 0.118$^{\arcsec}$/px (seeing limited), 0.13$^{\arcsec}$/px (from space),
 0.026$^{\arcsec}$/px.}
 \label{fig:resolution}
\end{figure*}
\end{centering}

\begin{centering}
\begin{figure*}
 \includegraphics[width=17cm]{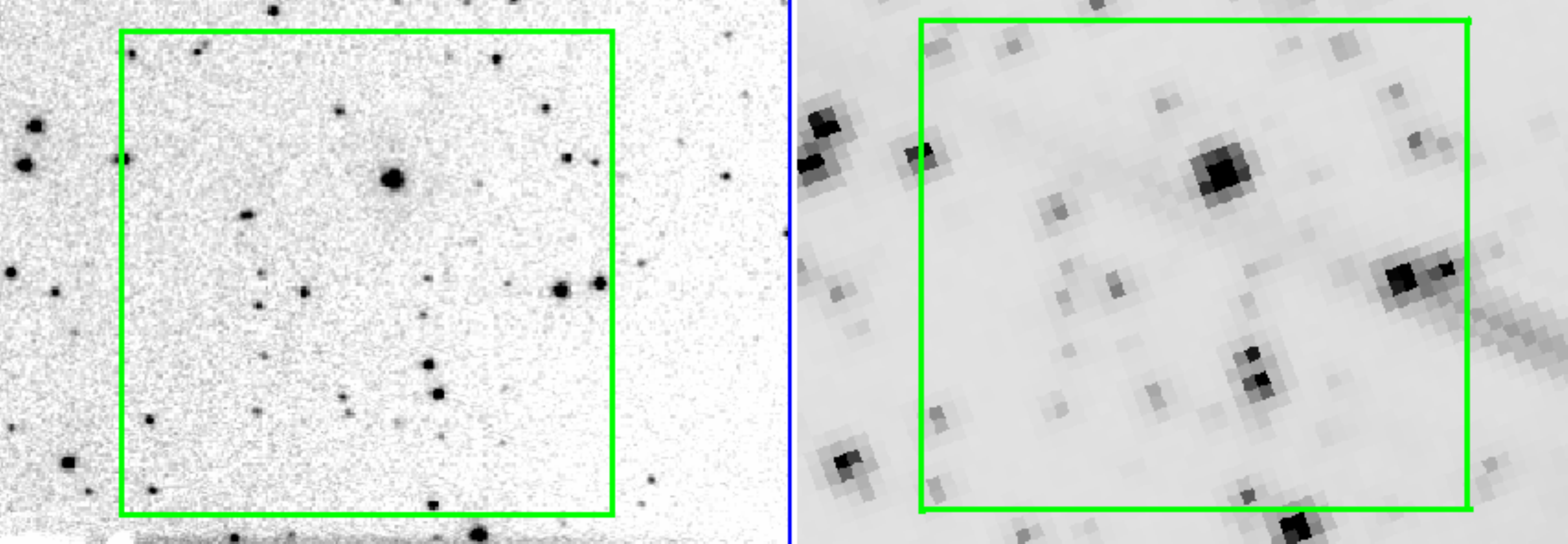}
 \caption{A zoom on the same PISCES (left) and WFC3 (right) images of Figure 
 \ref{fig:resolution}. The green square highlights the same sky region of 
 5$\arcsec\times$5$\arcsec$. The comparison clearly shows the gain in spatial 
 resolution of the PISCES data.}
 \label{fig:resolution2}
\end{figure*}
\end{centering}

Figure \ref{fig:map} shows an SDSS image of M\,15 with superimposed the
footprint of the adopted cameras: LUCI1 (black square), PISCES (blue), ACS
(magenta) and WFC3 (red).  The LUCI1 field (4$\arcmin\times$4$\arcmin$) was
selected such to include both the center of the cluster and the PISCES field,
which is located 2.7$\arcmin$ from the center. Note the small field-of-view
covered by the PISCES camera (21$\arcsec\times$21$\arcsec$). The HST/ACS field
is centered on M15, but due to the different position angle it is not overlapped
with the PISCES pointing, which on the other hand falls into the HST/WFC3 set.
 Figure \ref{fig:resolution} presents a comparison of the LBT/LUCI1 (left),
HST/WFC3 (center), and LBT/PISCES (right) region corresponding to the full
field-of-view of the PISCES camera. Other than the impressive improvement when
comparing data from the same telescope but without (left panel) and with (right
panel) assistance from the adaptive optics, it is clear that PISCES provide the
best spatial resolution also when compared to the WFC3 (central panel). It is
worth noting that the three images are stacked medians, thus the total exposure
time is different in the three cases. Moreover, while the left and right panels
show images in the $K_{\mathrm{s}}$, we only had available WFC3 images in the
$F160W$ filter, which is close to the $H$ band. To highlight even more the
PISCES performances, Figure \ref{fig:resolution2} shows a zoom of the PISCES
(left) and WFC3 (right) stacked images already shown  in Figure
\ref{fig:resolution}. The green squares mark a region of 
5$\arcsec\times$5$\arcsec$. The image discloses at first glance that the number
of sources is similar in both cases, suggesting a similar limiting magnitude (it
is worth recalling here the shorter total exposure time in both PISCES  $J$ and
$K_{\mathrm{s}}$, see Table 2). On the other hand, the contrast in the PISCES
images is by far better, and one can easily see that elongated sources in the
WFC3 field are well separated in the PISCES image, such as those next to the top
left corner of the green square.

 \begin{table}[ht!]
 \begin{center}
 \caption{Observations log.}
 \begin{tabular}{ccccc}
 \hline
 \hline
 \textit{Telescope} & \textit{Sensor} & \textit{filter} & \textit{Exposures}  & \textit{Total Time} \\
 \textit{ } & \textit{ } & \textit{ } & \textit{[s]}  & \textit{[s]} \\
  \hline
  LBT  & LUCI1  &  J & 13$\times$20 13$\times$40   & 780   \\
       &        &  K & 26$\times$40   & 1040   \\
       & PISCES &  J & 20$\times$30   &  600   \\
       &        &  K & 42$\times$15   &  630   \\
  HST  & WFC3   &  $F160W$   &  3$\times$200 6$\times$250   & 2100 \\
 \hline
 \hline
 \end{tabular}
 \end{center} 
 \label{tab:tab02}
 \end{table}


	\subsection{Data reduction and photometry}\label{sec:reduction}

The acquisition and the basic reduction have been performed following a
homogeneous  approach both for PISCES and LUCI1 data: raw images have been
secured by dithering the telescope within a 100 pixel random pattern to ensure
good removal of bad pixels. Single images have been dark-subtracted,
flat-fielded, resampled to remove geometrical distortions, and registered. For
the $K_{\mathrm{s}}$ filter, a superflat obtained with the off-source sky images
have been obtained and applied to the images to improve the low-frequency
flat-field removal.

The LUCI1 and WFC3 data have been independently reduced following the
prescriptions  of \citet{monelli10b}, and using a standard procedure based on
the  DAOPHOTIV/ALLSTAR/ALLFRAME suite of programs \citep{dao,alf}. Individual 
PSFs have been modeled for each image, using semi-automatic routines. The input
list of stars for ALLFRAME was generated registering and matching the individual
catalogues from single images.

The case of the PISCES data deserved particular attention, because adaptive
optics  may provide PSFs characterized by spatial variations across the  field
due to anisoplanatism. This is especially true for the shorter wavelength
$J$-band images, while the  PSF in the $K_{\mathrm{s}}$ is typically more stable
even at the largest distances from the guide stars, as it scales with the
isoplanatic angle and progressively with the wavelength. Therefore, to
perform the photometry on these images we adopted the ROMAFOT  suite of programs
\citep{buonanno83, buonanno89b}. The  PSF photometry with ROMAFOT is more
lengthy when compared with similar packages  available in the literature.
However, it is has the key advantage of a graphical  interface that allows the
user to improve the local deconvolution of stellar  profiles. The latest version
of the code (Ferraro et al. 2015, in preparation)  has been optimized to perform
accurate photometry of crowded stellar fields  on images collected with AO
systems. In particular, it takes into  account the spatial variation of
the PSF across the field of view and the  variation of the asymmetric, egg-like
shape of the PSF. A preliminary discussion  of the numerical algorithms and of
the approach adopted to deal with the quoted  images has already been presented
in \citet{fiorentino14b}. In passing we note that photometry performed using
asymmetric PSFs on ground-based (LUCI1@LBT) and  space (WFC3@HST) images gives
magnitudes that are, within the errors, identical to those measured using
other photometric packages.

The final adopted photometry was obtained in two steps. First, ROMAFOT was run
on the mean $J$ and $K_{\mathrm{s}}$-band images in order to create the master
list of objects. The final photometry was obtained reducing the individual
images, and averaging the derived magnitudes. A Moffat analytic function was
adopted to model the PSF, with $\beta$=2.0, 2.5 and $\sigma$=2.70, 2.05 for the
$J$ and  $K_{\mathrm{s}}$ images, respectively.

The final photometric catalogues were calibrated into the 2MASS photometric 
system in two steps. First, the LUCI1 photometric catalogue was calibrated using 
$\sim$200 stars in common with 2MASS and covering the entire field of view. Then 
three dozen local standards of the LUCI1 catalogue were used to calibrate the 
photometric catalogue based on PISCES images. In this context it is worth 
mentioning the crucial role that faint local standards play in the accurate 
calibration of images collected with AO systems. The use of NGS for the tip 
tilt correction and the modest field of view of current AO systems implies 
the selection of crowded stellar fields. In these cluster regions the 
photometric quality of the 2MASS local standards is quite poor, moreover, 
even the faintest 2MASS stars are in these regions saturated. These are the 
reasons why the calibration of NIR images collected with AO systems does 
require a double step in the calibration using 4-8m class telescopes to 
improve the limiting magnitude and the photometric accuracy of local standards 
\citep{bono10a}.

Keeping in mind the above caveats, Figure~\ref{fig:calibration} shows the 
approach we adopted to perform the absolute calibration. From left to right, 
the three rows show the residuals of the calibration as a function of 
magnitude (top: $J$; bottom: $K_{\mathrm{s}}$), the X and the Y coordinate. 
Small dots indicate the 2MASS stars in common with LUCI1, while the open 
squares show the $\sim$35 LUCI1 stars used as local standard to calibrate 
the PISCES catalogue. No apparent trends are visible, and the residuals have 
null mean.

\section{Results}\label{sec:cmds}

Figure \ref{fig:cmd} shows NIR CMDs of the central regions of M\,15 based on
four  different imagers. From left to right, the ($K_{\mathrm{s}}$,
$J-K_\mathrm{s}$) CMD based on: a) LBT/LUCI1 covering a field of view of 16
arcmin squared ($\sim$17,000 stars); b) the intersection of space $F160W$-band
(HST/WFC3) and ground-based  $K_{\mathrm{s}}$-band (LBT/PISCES) photometry
covering a field-of-view (FoV) of 0.19 arcmin squared ($\sim$380 stars);  c)
LBT/PISCES covering a FoV of 0.19 arcmin squared ($\sim$450 stars). The jump in limiting magnitude is compelling when
moving from seeing--limited (LBT/LUCI1, b)) to adaptive optics (LBT/PISCES, c))
NIR images collected with the same telescope. Indeed, the use of PISCES camera
together with the FLAO \citet{esposito12, riccardi10} allows us to move the
limiting magnitude in the K-band down to $\approx$22.5--23.0 mag. To our
knowledge the deepest $K_{\mathrm{s}}$-band photometry ever performed in a GC.  

\begin{centering}
\begin{figure*}
 \includegraphics[width=17cm]{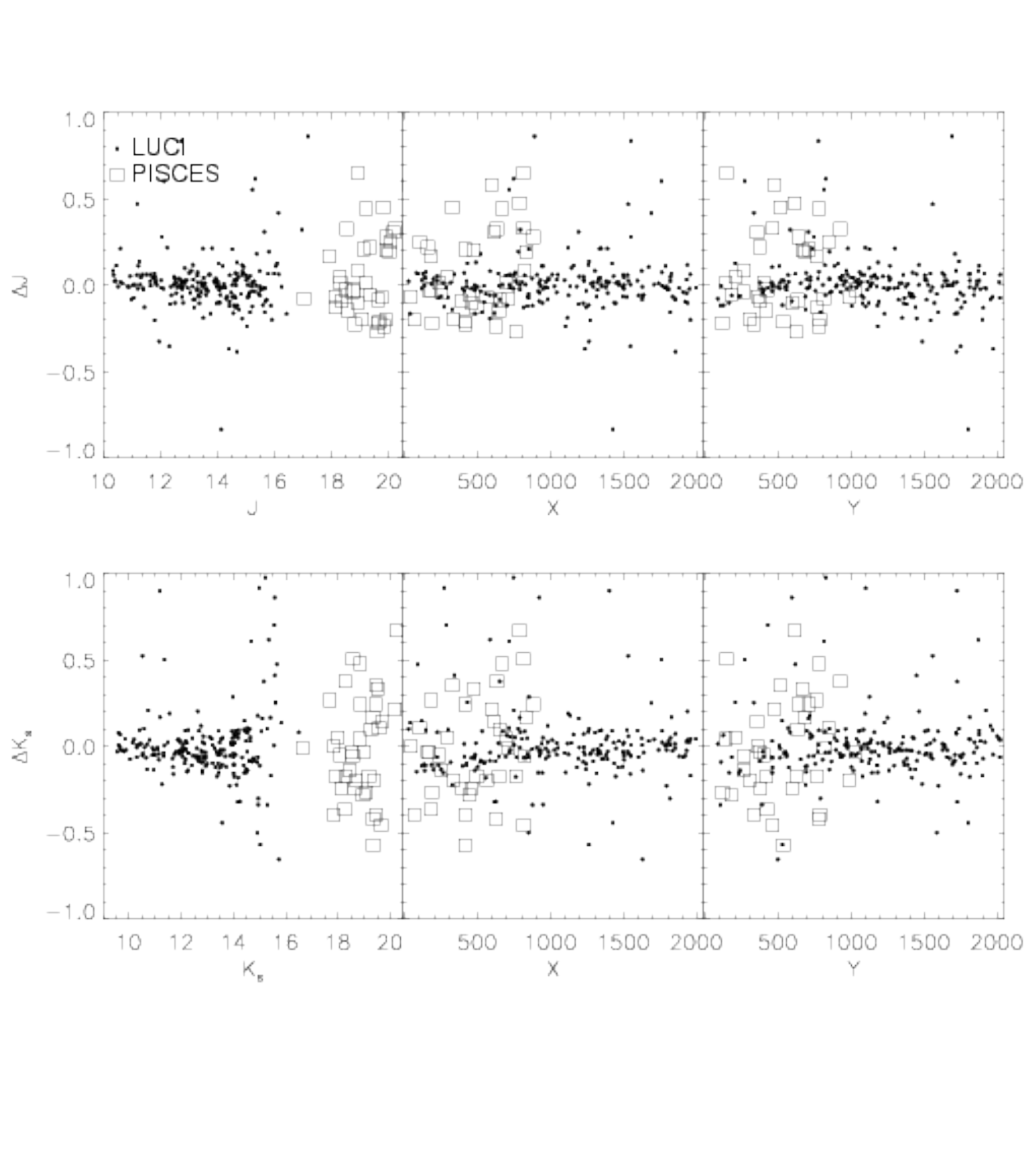}
 \vspace{-2cm}
 \caption{Residuals of the calibrations as a function of magnitude ({\it 
 Left}), X coordinate {\itshape Center}, and Y coordinate {\itshape Right}. 
 top and bottom rows refer to the $J$ and $K_{\mathrm{s}}$ bands,
 respectively. Large dots show the more than 200 stars in common between the LUCI
 catalogue and 2MASS. Big open squares compare the $\sim$35 stars in
 common between LUCI1 and PISCES, at significantly fainter magnitude. Note that
 the range of X and Y are different because of the size of the two cameras.}
 \label{fig:calibration}
\end{figure*}
\end{centering}

\begin{centering}
\begin{figure*}
 \includegraphics[width=16cm]{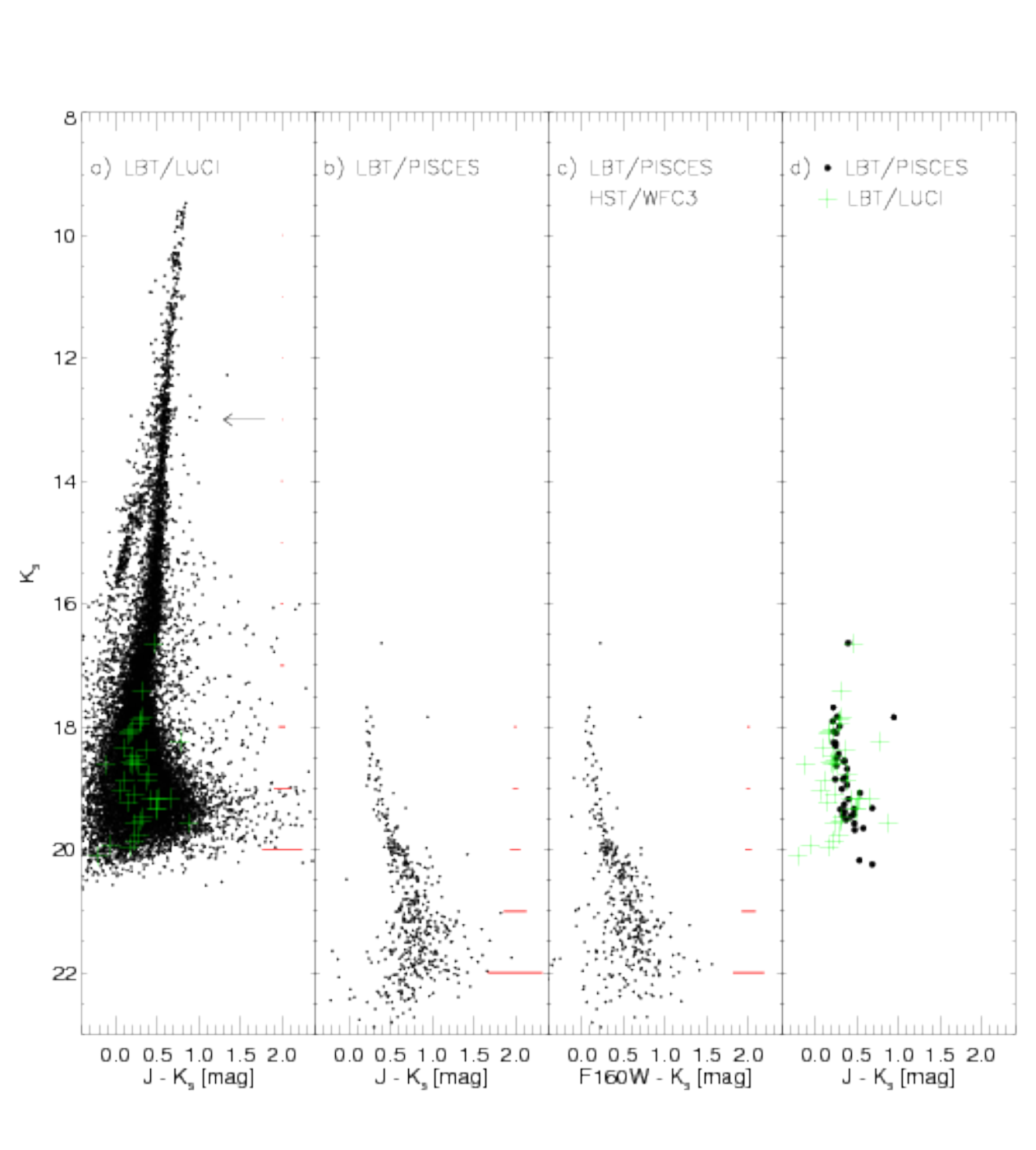}
 \caption{$(K_{\mathrm{s}},J-K_{\mathrm{s}})$ CMDs of M\,15. {\it Panel a) - }
 Data from LBT/LUCI1. The horizontal arrow marks the position of the 
 RGB bump. The green pluses show the stars in common with the PISCES photometry.
 {\it Panel b) - } LBT/PISCES CMD of the outer M\,15 field.
 {\it Panel c) - } Mixed CMD with HST/WFC3
 [F160W] together with LBT/LUCI1 [$K_{\mathrm{s}}$] data.
 {\it Panel d) - } Comparison between the CMD of the stars in common between 
 LUCI1 (green pluses sa as in panel a)) and PISCES (large black dots).}
 \label{fig:cmd}
\end{figure*}
\end{centering}

Panel d) shows a direct comparison between the CMD of the stars in
common with PISCES (large black dots) and LUCI1 (green pluses). The two 
overlap over a magnitude range of $\approx$ 3.5 magnitudes, from 
$K_{\mathrm{s}}\approx$16.6 mag to $K_{\mathrm{s}}\approx$20.1 mag. 
The main sequence based on PISCES images is narrower than the LUCI1 one, 
and no systematics appear in the comparison (see also
Figure~\ref{fig:calibration}). Interestingly enough, the sequence of green 
symbols plotted in panel a) discloses a smaller photometric dispersion than 
the bulk of the main sequence stars in the LUCI1 field. The difference is 
mainly due to the low crowding level of these cluster regions. Indeed, 
the 450 stars measured in the PISCES FoV imply a density of 0.67
stars arcsec$^{-2}$ sq., that is one star every $\sim$2,330 px$^2$. 
This context is quite different when compared with the central pointing, 
since in this region the typical stellar density is a factor of $\approx$50
higher.

The CMD based on LUCI1 images allows the identification of the typical 
evolutionary features of a GC. It covers  more than 10 magnitudes in  the
$K_{\mathrm{s}}$ band, and ranges from the tip of the RGB
($K_{\mathrm{s}}\sim$9.5 mag) down  to $\sim$2 mag below the main sequence
turn-off  ($K_{\mathrm{s}}\sim$18, $J-K_{\mathrm{s}}\sim$0.25 mag).
Moreover,  M\,15 also shows a well--populated horizontal branch ranging from
$K_{\mathrm{s}}\sim$16.5 to $K_{\mathrm{s}}\sim$14 mag with the slope
typical  of NIR CMDs. Finally, the RGB bump appears clearly in the luminosity
function of the RGB at magnitude $K_{\mathrm{s}}=13.00\pm0.05$ mag, indicated
by the arrow.  

In spite of the good quality of LUCI1 photometry, it is thanks to PISCES  and to
the FLAO system that we have been able to identify, for the first  time, the MSK
($K_{\mathrm{s}}\sim$21.5 mag) in a very metal-poor GC.  The quality of the
CMD based on LBT/PISCES is further supported by the  comparison with the CMD
based on both space and ground--based NIR images.    Data plotted in panel d) of
Fig.~\ref{fig:cmd} show that the limiting magnitude in  F160W is similar to the limiting
magnitude in the $J$-band. However,  the intrinsic error at fixed magnitude
seems larger in $F160W$ than in $J$-band CMD, and indeed the MSK
cannot be easily identified  in the above CMD. In passing we also note that the
exposure time in  F160W is 3.5 times larger than the exposure time in the
$J$-band.

\begin{centering}
\begin{figure*}
 \includegraphics[width=16cm]{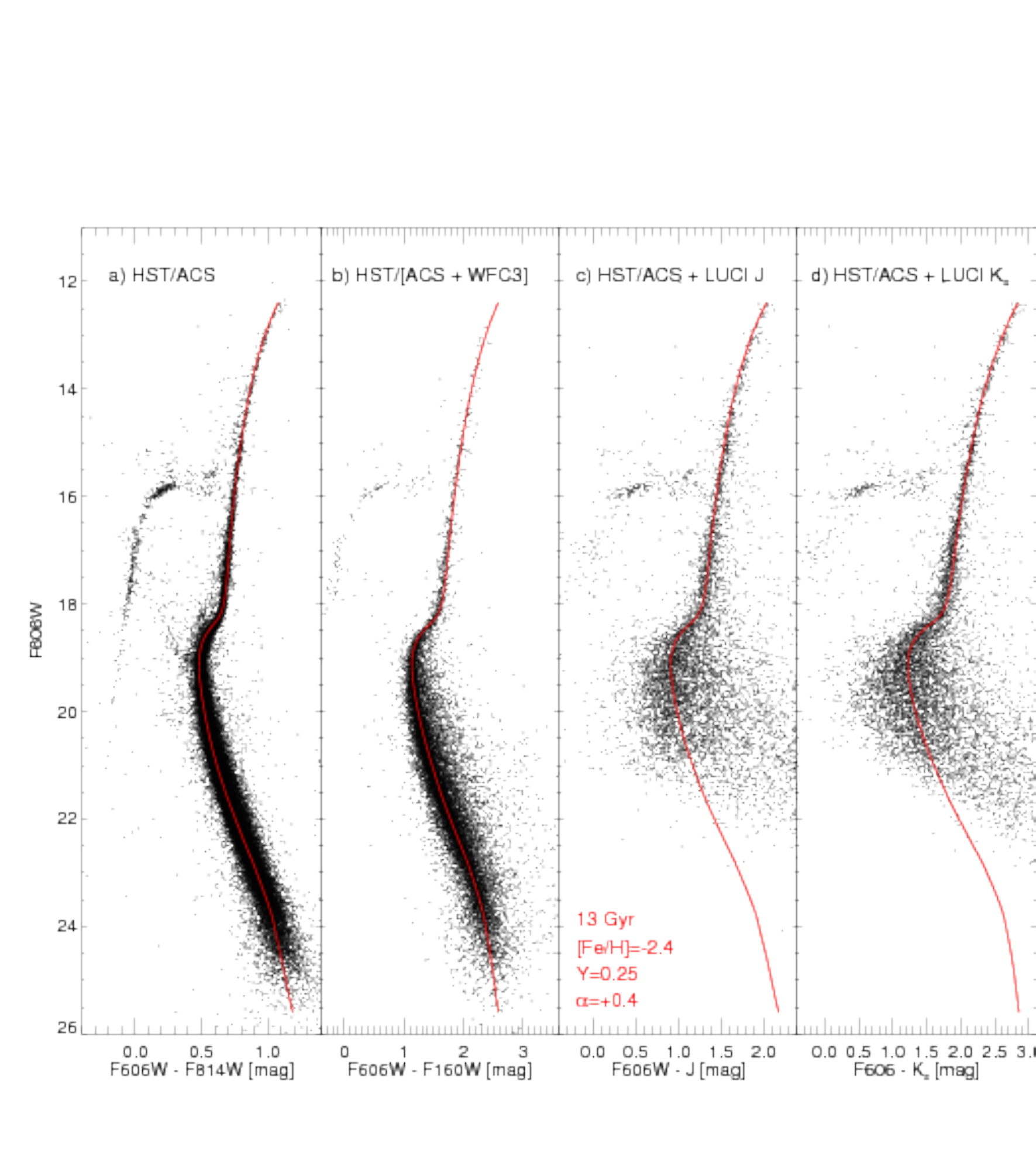}
 \caption{Optical and NIR CMDS based on HST data with superimposed the same 
 isochrone for the labelled parameters.}
 \label{fig:hst}
\end{figure*}
\end{centering}

\begin{centering}
\begin{figure}
 \includegraphics[width=8cm]{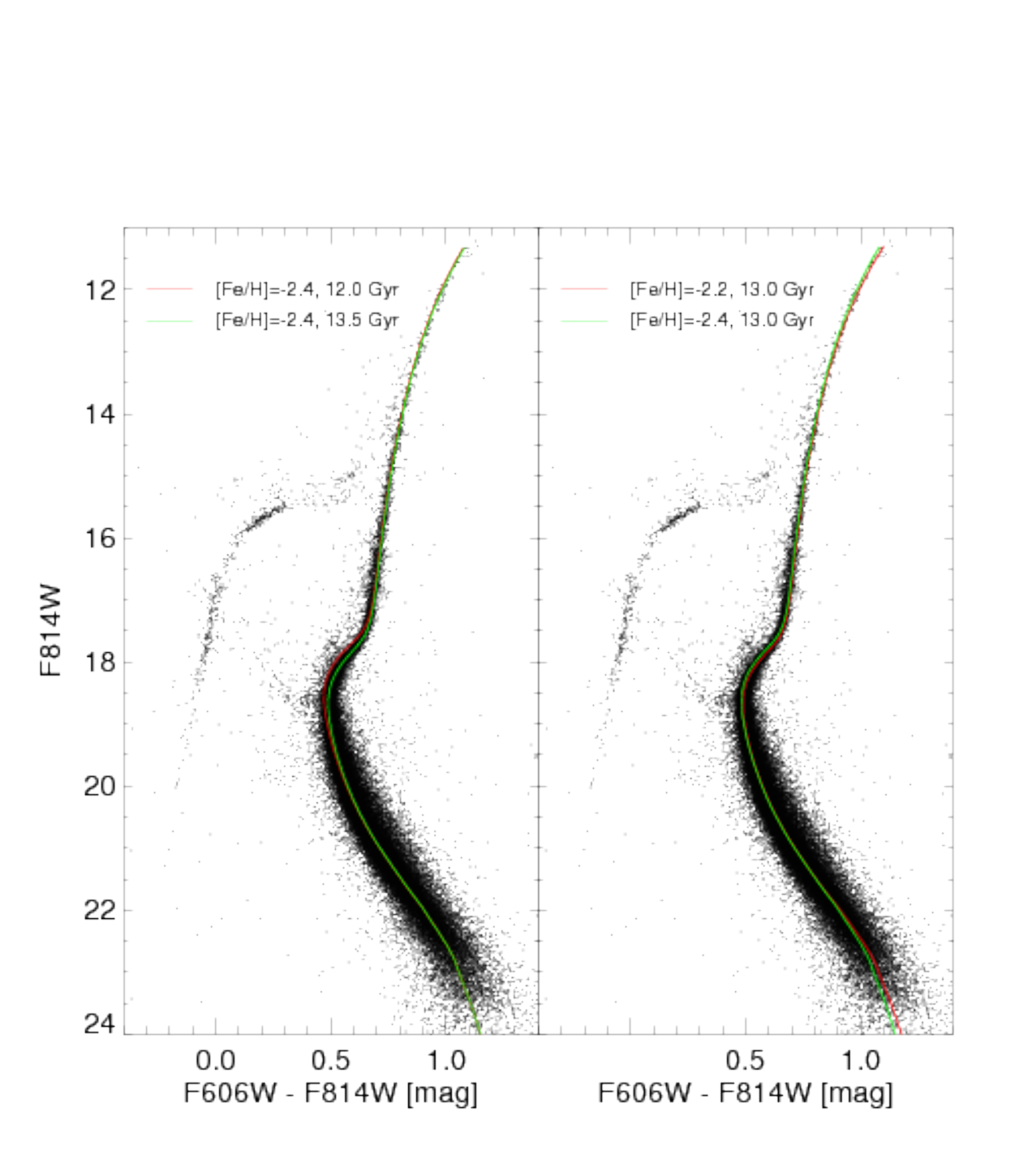}
 \caption{Comparison with isochrones of different ages ({\textit left}) and
 metallicities ({\textit right}). }
 \label{fig:agemetal}
\end{figure}
\end{centering}

We are dealing with photometric catalogues that have been collected using 
different telescopes equipped with different imagers and different sets of 
filters. Their possible systematics might affect the absolute age estimates.  
To constrain this effect, Figure \ref{fig:hst} and Figure \ref{fig:agemetal}
show the comparison of selected cluster isochrones with our data. A glance 
at the data plotted in this figure clearly shows the advantage of using 
CMDs based on optical and NIR photometric data. Indeed, the optical-NIR 
CMDs do cover a range in color that is at least a factor of two larger 
compared with the optical ones.  From left to right, Figure
\ref{fig:hst}  shows: a) the optical CMD based on ACS images ($F606W$,
$F606W$-$F814W$);  b) the optical--NIR CMD based on ACS and on WFC3 images
($F606W$, $F606W$-$F160W$); c) the optical--NIR CMD based on ACS and on LUCI1
$J$-band images ($F606W$, $F606W$-$J$); d) the optical--NIR CMD based on ACS
and on LUCI1 $K$-band images ($F606W$, $F606W$-$K_{\mathrm{s}}$).  The
cluster isochrone was computed by adopting the evolutionary tracks provided
by \citet{vandenberg14b}. The isochrones were transformed into the 
observational plane by adopting the color-temperature relation by
\citet{casagrande14}. We adopted an iron abundance of [Fe/H]=-2.4
\citep{kraft03}  an $\alpha$-enhancement of $\alpha$=+0.4
\citep{sneden97,sneden00} and primordial  helium content of $Y$=0.25, and a
cluster age of 13 Gyr (red line). Current  theoretical framework is fully
consistent with the set of isochrones  adopted by \citet{bono10a}.  We performed a series of test to constrain the optimal true distance 
modulus and reddening that provide a good simultaneous agreement between 
theory and observations in the four CMDs plotted in Fig. 5. We found, 
using the Cardelli et al. (1989) reddening law, that a cluster 
reddening of E(B-V)=0.08 mag together with a distance modulus of 
$(m-M)_0$=15.14 mag \citep[][,2010 edition]{harris96} do provide 
a good agreement in the quoted optical-NIR CMDs. 
The cluster reddening agrees with the value suggested by  \citet{sandage81}, 
but marginally smaller than the more recent estimates by \citet{schlafly11} 
(E(B-V)=0.10). The true distance modulus agrees, within the errors,
with different estimates available in the literature 
\citep{durrell93,vandenbosch06}. Note that the assumption of a larger 
extinction would imply, at fixed age, a systematic drift of the isochrones 
toward redder colors. The new discrepancy could be alleviated by a decrease in 
cluster age, but younger isochrones are characterized by a slope of the 
subgiant branch (SGB) that is too steep when compared with the data. This 
difference becomes more evident in the optical planes where the SGB is 
remarkably thin. The above evidence 
indicates that adopted distance and reddening are mainly constrained by the 
morphology of SGB and RGB. The former one playing a crucial role, since it 
is less affected by uncertainties in the adopted mixing length 
\citep{salaris15}.

 The CMDs plotted in panels {\itshape c)} and {\itshape d)} also suggest 
a good agreement between theory and observations. Indeed, only a marginal 
shift in color is present at the base of the RGB in the $F606W$,$F606W - J$ 
CMD. The anonymous referee also noted a similar shift in the RGB region 
between the HB and the base of the RGB. In the quoted cases, the isochrone 
are once again marginally redder than the observed RG stars. The above 
empirical evidence brings forward two relevant points. 

{\em i)}-- Theory and observations disclose an overall very good agreement 
over more than ten magnitudes. The agreement becomes even more compelling 
if we take account of the fact that we are dealing with optical and NIR data 
collected with space and ground-based facilities. This finding also supports 
the adopted chemical composition, the bolometric corrections and the 
color-temperature transformations together with the adopted true distance 
modulus and cluster reddening. Similar results have also been obtained in 
the literature by different groups \citep{dotter08,sarajedini09a,bressan12,
dellomodarme12,vandenberg14b} thus further supporting the current accuracy 
era of stellar astrophysics.   

{\em ii)}-- The difference in color at the level of 0.03 mag between 
theory and observations has a marginal impact on the methods we are using 
to estimate the absolute cluster age. Indeed, they rely either on absolute 
(MSTO) or on relative magnitudes (MSK).

Fig \ref{fig:agemetal} shows a comparison with isochrones for different 
assumptions on the $\alpha$ enhancement and the metal content. Left panel 
presents two isochrones of 12 (red line) and 13.5 Gyr (green). The effect of 
age appears clearly in the MSTO region, but as expected it does not affect 
neither the RGB nor the low main sequence. The right panel present two 
isochrones of 13 Gyr, with [Fe/H]=-2.2, -2.4 (red and green line, respectively) 
for different age and metallicity assumptions.

\section{The absolute age of M\,15}\label{sec:age}

To estimate the absolute age of M\,15 we devise here a double approach. The
first is based on the MSTO position, and the second on the magnitude difference
between the MSTO and the main MSK Table 3 summarizes the observables and
the cluster ages based on the two quoted methods. From left to right the
columns give the adopted CMD, the imager, the apparent magnitude, the color
index of the MSTO (m$_{MSTO}$, CI$_{MSTO}$), and of the MSK (m$_{MSK}$,
CI$_{MSK}$) together with the two cluster ages.

	\subsection{The TO method}

The position of the MSTO is determined in the observational planes the bluest MS
point of the ridge line. Similarly, we estimate the magnitude and color index
of the MSTO for a set of four isochrones of fixed  metallicity ([Fe/H] = $-$2.4,
$\alpha$ = +0.4, $Y$ = 0.25) and age between 12.0 and 13.5 Gyr (see \S
\ref{sec:cmds}) once rescaled for the proper distance and reddening. This allows
us to determine that, at least in this age range, a linear relation exists
between the MSTO magnitude and the age.  The slopes of the above relations
were obtained with a linear Least Squares fit. The coefficients are listed in
Table 4, together with the predicted  magnitudes of the MSTO and of the MSK as a
function of cluster age. These slopes are a measure of the sensitivity of
the MSTO as an age indicator in the different bands. For example, we derive that
the $F606W$ band is $\sim$1.8 times more accurate than the $K_{\mathrm{s}}$
band.

The age corresponding to M\,15 is derived by interpolating the previous
relations assuming the observed MSTO magnitude. The error budget has to take
into account various sources, including error on the photometric
calibration, the MSTO magnitude, the reddening and distance\footnote{The
error in the age is totally dominated by the propagated error in the distance
estimate. We note that the distance value from \citet{vandenbosch06}
provides error a factor of 2 smaller, which would imply a reduction of 
$\sim$30\% on the error on the absolute age determined with the MSTO.}. In the case of the
present data set, the photometric error varies depending on the filter used,
from $\sim$0.011 in the case of the $F814W$ filter to $\sim$0.04 mag for the
LUCI1 $K_{\mathrm{s}}$. Nevertheless, while the differential reddening is
almost negligible along the line of sight of M\,15 (E(B-V)=0.08$\pm$0.01 mag,
\citealt{sandage81}), the dominant source of error is the distance ($\pm$0.15
mag, \citealt{durrell93}). We derive an absolute age ranging from
12.8$\pm$2.0 ($F814W$ band) to 14.0$\pm$3.0 Gyr ($K_{\mathrm{s}}$), with
weighted mean value of 12.9$\pm$2.6 Gyr.

	\subsection{The MSK method}

The second approach is based on the magnitude difference between the MSTO and
the MSK, $\Delta M({\rm MSTO-MSK})$. This method is based on the fact the
magnitude and color of MSK in the low-mass regime of the MS are, at fixed
chemical composition, essentially independent of cluster age \citep{bono10a}.
The key advantage of the MSK is that it 
is caused by collision-induced absorption (CIA) opacities of both 
H$_2$--H$_2$ and H$_2$--He in the surface of cool dwarfs \citep{borysow01,
borysow02}. The MSK is independent of cluster age and 
anchored in a region of the MS that is marginally affected by uncertainties 
in the treatment of the convective regime that is nearly adiabatic 
\citep{saumon08}. Recent empirical evidence indicates that the error in 
the absolute age of GCs based on this method are on average a factor of two 
smaller than the canonical ones \citep{bono10a, dicecco15}.      

Therefore, this method offers a powerful observable to constrain the cluster
age either as a color or as a magnitude difference between the bend and the
cluster MSTO, using both optical and infrared filter combinations. In the case
of the present data set, we determined the position of both points using the
ridge line of the cluster in the different CMDs \citep[see e.g.][]{dicecco15}. 
Note that lack of a sizable sample of MSTO stars in the PISCES photometry 
is a direct consequence of the modest field of view covered by the camera 
and by the radial distance of the pointing. To constrain on a more quantitative 
basis this relevant issue we selected the same cluster region covered by 
PISCES in the LUCI1 photometry and we found that it only includes 35 stars 
and the bulk of them are located at magnitudes fainter than the MSTO (see 
panel d) of Figure \ref{fig:cmd}).  

\begin{centering}
\begin{figure}
 \includegraphics[width=8cm]{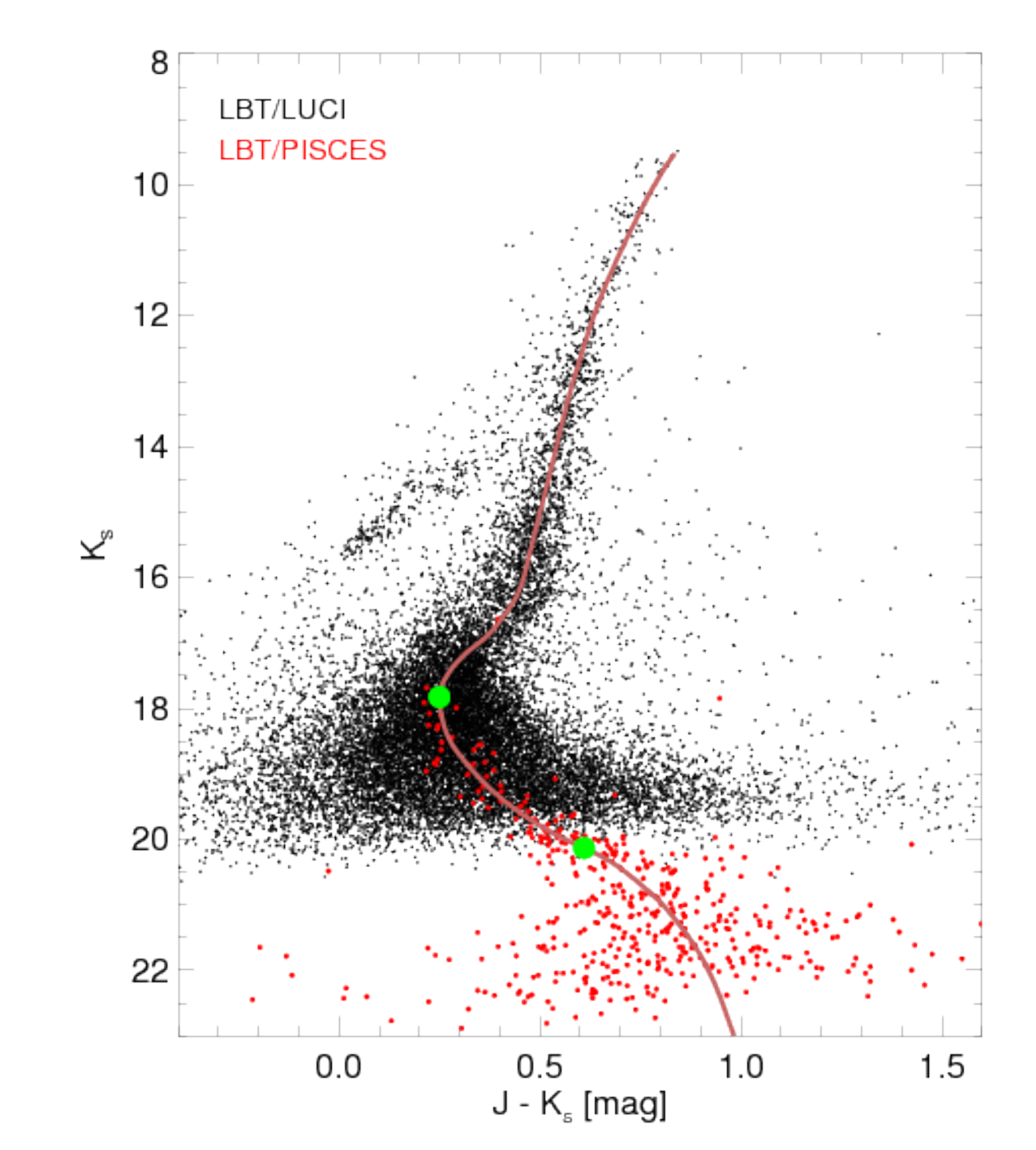}
 \caption{$(K,J-K)$ CMDs from LBT data (black: LUCI1; red: PISCES) with superimposed
 the derived ridge line. The big dots mark the MSTO and MSK points.  }
 \label{fig:ridge}
\end{figure}
\end{centering}

To overcome this issue, we merged the LUCI1 and PISCES CMDs, and derived a
unique ridge line. In this way, the TO region is sampled by the large number of
stars in the LUCI1 photometry, while the MSK is sampled by the PISCES deep
catalogue. In particular, following \citet{bono10a}, we  define the color and
magnitude of the MSK at the maximum curvature point in the low part of the MS. 
The outcome is shown in Figure \ref{fig:ridge}, where the combined LUCI1 and
PISCES CMD are shown, together with the cluster ridge line, the MSTO and MSK.
The approach adopted to compute the cluster ridge line has  been discussed in
detail by \citet{dicecco15}. The corresponding theoretical  points were estimated
using the same approach on the same isochrones  adopted to estimate the cluster
absolute age with the canonical MSTO  method.

Similarly to the age determination based on the MSTO, we derive linear relations
between $\Delta M({\rm MSTO-MSK})$ and the age (see Table 4). In this case, the
method is slightly more sensitive than adopting the absolute value of the MSTO,
by $\sim$ 15\%. The age is therefore  derived interpolating the empirical value
derived from the CMDs, and is reported in the last columns of Table 3. The
estimated age ranges from 12.9$\pm$0.9 ($F606W$ band) to 13.7$\pm$1.0 Gyr
($K_{\mathrm{s}}$),  with mean value 13.3$\pm$1.1 Gyr.

	\subsection{Comparison with literature values}

 The values shown in the last two columns Table 3 disclose a general
good agreement between the two derived age values. The estimate derived
with the ($K_{\mathrm{s}}, J-K_{\mathrm{s}}$) CMD is marginally
higher than the other ones, but still well within the error bars. Also,
the age derived with the MSTO are 0.3 to 0.9 Gyr smaller than the corresponding
value derived with the MSK approach.

The age of M\,15 has been subject to a large number of investigations
\citep[e.g.\ ][]{salaris97, salaris98,salaris02, mcnamara04,deangeli05}. We
stress that a straight comparison of literature estimates is complicated by
the different theoretical scenarios adopted. However, the age derived here
is in good agreement with recent estimates available in the literature
(12.8$\pm$0.6 Gyr, \citealt{marinfranch09}; 12.75$\pm$0.25$\pm$1.5 Gyr,
\citealt{vandenberg13}). Notably, these estimates are based on a 
difference approach. The former analysis use relative age of a population
of clusters, anchored to an absolute scale using clusters with
well-determined distance (NGC\,6752 via the subdwarf-based method).  The
latter uses the absolute magnitude of the MSTO. Note that
we are using the same set of optical data as \citet{vandenberg13}, and
indeed we independently obtain a perfectly consistent value of the age
based on the MSTO luminosity, close to 12.8 Gyr.

\section{Discussion}\label{sec:discussion}

	\subsection{Comparison with NGC\,3201 and metallicity dependence}

In the present work we studied NIR photometry of the very metal-poor
cluster M\,15. Nonetheless, it is interesting to extend the analysis to a
more metal-rich regime in order to explore the dependency of the $\Delta
M({\rm MSTO-MSK})$ method on both the age and the metallicity of the target
system.  \citet{bono10a} presented a similar analysis for the cluster
NGC\,3201 ([Fe/H] $\sim -$1.5). By adopting the same theoretical framework
applied here we derive analogous linear relations correlating the magnitude
difference $\Delta M({\rm MSTO-MSK})$ with age, assuming [Fe/H] = $-$1.5.
The obtained derivatives (e.g.: 0.13$\pm$0.01 mag Gyr$^{-1}$ and
0.08$\pm$0.01 mag Gyr$^{-1}$ for the $F606W$ and the  $K_{\mathrm{s}}$
band, respectively) are very similar to those derived for more metal-poor
isochrones suitable for M\,15. This supports that the  $\Delta M({\rm
MSTO-MSK})$ diagnostic can be fruitfully used over a wide range of
metallicities.


 \begin{table*}[ht!]
 \begin{center}
 \caption{Age determinations}
 \begin{tabular}{lccccccccc}
 \hline
 \hline
 \textit{CMD} &     \textit{Cameras}   &  \textit{m$_{MSTO}$} & \textit{CI$_{MSTO}$} & \textit{m$_{MSK}$} & \textit{CI$_{MSK}$} &   \textit{t$_{MSTO}$} &   \textit{t$_{{\rm MSTO-MSK}}$} \\
  \hline
$F814W, F606W - F814W$                & ACS       &  18.801$\pm$0.011   &  0.490$\pm$0.016    &   21.642$\pm$0.036	 &  0.829$\pm$0.052   &      12.8$\pm$2.0   &   13.3$\pm$0.6     \\
$F160W, F814W - F160W$                & WFC3+ACS  &  18.027$\pm$0.040   &  0.652$\pm$0.041    &   20.296$\pm$0.051	 &  1.045$\pm$0.059   &      12.5$\pm$2.7   &   13.4$\pm$1.3     \\
$F160W, F606W - F160W$                & WFC3+ACS  &  18.027$\pm$0.040   &  1.147$\pm$0.041    &   20.377$\pm$0.051	 &  1.838$\pm$0.061   &      12.6$\pm$2.7   &   12.9$\pm$1.3     \\
$K_{\mathrm{s}}, J - K_{\mathrm{s}}$  & PISCES    &  18.010$\pm$0.043   &  0.238$\pm$0.044    &   20.160$\pm$0.290	 &  0.581$\pm$0.299   &      14.0$\pm$3.1   &   13.7$\pm$1.4     \\
 \hline
 \end{tabular}
 \end{center} 
 \label{tab:tab03}
 \end{table*}


Moreover, by comparing the absolute position of the MSK in different
isochrones, we find that: {\em i)} at fixed metallicity, the magnitude of
the MSK changes by at most 0.02 mag, for ages larger than 10.5 Gyr; {\em
ii)} similarly, at fixed age the MSK moves by $\sim$0.02 mag, when moving
from [Fe/H] = $-$2.4 to [Fe/H] = $-$1.5. Interestingly, these values seem
independent of the wavelength, at least in the spectral range covered by
the $F606W$ to $K_{\mathrm{s}}$ the passbands. Overall, the position of the
MSK seems to be a reliable anchor,  marginally dependent on the age and the
metallicity, at least in the ranges investigated so far. However, a more
systematic theoretical investigation is needed, in order to to constrain in
detail the sensitivity of the MSK, and in turn of the $\Delta M({\rm
MSTO-MSK})$ method, over the full range of ages and metallicities typical
of Galactic GCs.

 \begin{table}[t]
 \begin{center}
 \caption{Magnitude v$_s$ ages derivatives.}
 \begin{tabular}{lcc}
 \hline
 \hline
 \textit{Mag} & \textit{dMag(TO)/dt} & \textit{dMag(TO-MSK)/dt} \\
  \hline
$F606W$                &  0.09$\pm$0.01  &  0.11$\pm$0.01  \\
$F814W$                &  0.08$\pm$0.01  &  0.09$\pm$0.01  \\
$J$                    &  0.06$\pm$0.02  &  0.07$\pm$0.02  \\
$F160W$                &  0.06$\pm$0.01  &  0.07$\pm$0.01   \\
$K_{\mathrm{s}}$       &  0.05$\pm$0.02  &  0.06$\pm$0.02   \\
\hline
\textit{Mag} & \textit{dCI(TO)/dt} & \textit{dCI(TO-MSK)/dt} \\
\hline
$F606W$-$F814W$       &  0.01$\pm$0.01  &  0.02$\pm$0.01  \\
$F606W$-$F160W$       &  0.03$\pm$0.01  &  0.02$\pm$0.01  \\
$J$-$K_{\mathrm{s}}$  &  0.01$\pm$0.01  &  0.01$\pm$0.01   \\
 \hline
 \end{tabular}
 \end{center} 
 \label{tab:tab04}
\vspace{0.3cm}
 \end{table}


		\subsection{Cosmological implications}

There is mounting evidence that we are in the era of precision cosmology. 
Recent estimates of the Hubble constant suggest a precision of the order of 
3\% (74.3$\pm$2.1 km s$^{-1}$ Mpc$^{-1}$, \citealt{freedman12};  73.8$\pm$2.4
km s$^{-1}$ Mpc$^{-1}$, \citealt{riess11}).  Plain physical  arguments
suggest, that in a flat Universe, the age of the  Universe --$t_0$-- is
connected with the Hubble constant --$H_0$--, the  matter density parameter
--$\Omega_m$-- and with the dark energy  density --$\Omega_\Lambda$-- by the
following relation \citep{dekel97}:

\begin{equation}
 t_0 = [ 1 - (\Omega_m - 0.7\Omega_\Lambda)/5.8 ] / (1.3\times h_0)    
\end{equation}

\noindent where $h_0$ = $H_0$/100 km s$^{-1}$Mpc$^{-1}$ is the current 
espansion rate of the universe and $t_0$ is the age of the  Universe today in
units of 10$^{10}$ Gyr. 

Using the recent estimates of the cosmological parameters provided  by WMAP
and Planck ($\Omega_m$=$0.315^{+0.016}_{-0.018}$, 
$\Omega_\Lambda$=$0.685^{+0.018}_{-0.016}$, \citealt{hinshaw13,planck14}) we
found  $t_0$ = 10.70$\pm$0.82 Gyr\footnote{Note that we did not assume  an
Eisten-deSitter cosmological model ($\Omega_m$=1, $\Omega_\Lambda$=0) 
because the age of the Universe we obtain is systematically younger 
($t_0$=8.6$\pm$0.28 Gyr) than suggested by recent ``stellar'' and 
``cosmological'' estimates.}. The above cosmological age is within  1$\sigma$
of current stellar ages. However, it is on the short  limit if we also take
account of the time for structure formation  (z$\sim$8, t$<$ 1 Gyr).        A
similar approach to constrain the age of the Universe today  is to use
together with the above cosmological parameters also  the estimate of $H_0$
provided by Planck+WMAP  ($H_0 =$67.3$\pm$1.2 km s$^{-1}$Mpc$^{-1}$). It is
worth mentioning  that $H_0$ is a prior in the CMB solution ranging from 20
to 100 km s$^{-1}$Mpc$^{-1}$. Note that in this context the precision on
$H_0$ is of the order  of 1.8\%. The age of the Universe we found is
$t_0$=11.75$\pm$0.21 Gyr, while the detailed inversion of the CMB maps
provides $t_0$=13.817$\pm$0.048 Gyr. The above
estimates appear in quite good  agreement with the current absolute ages of
oldest GCs, and we note that a smaller age of 10.71$\pm$0.50 Gyr is derived
assuming the $H_0$ value from \citealt{riess11}.

However, we are facing the evidence that the uncertainties on cosmological
ages are systematically smaller than 1 Gyr for  estimates based on $H_0$ and
become smaller than a few hundreds  of Myrs for CMB determinations.  On the
other hand, the uncertainties affecting the absolute age  of GCs ranges from
2 (MSTO) to 1 (MSK) Gyr. The difference in  precision between cosmological and
stellar ages is going to  become even more prominent, since the next
generation of  experiments \citep{riess11,freedman12} plans to  improve by a
factor of two the precision on $H_0$. Taken at face  values the ``stellar''
estimates do not allow us to validate the  evaluations provided by
cosmology.  It is clear that accurate absolute ages for a sizable sample of
extremely  metal--poor GCs can shed new lights on this long-standing  and
intriguing problem.

\section{Conclusions and future remarks}\label{sec:conclusions}

We have presented new NIR data of the metal-poor globular cluster M\,15 obtained
with the LUCI1 and PISCES cameras available at the LBT, and complemented with
archival optical and NIR HST data. The analysis of the data raised important
points: {\em i)} ground-based adaptive optics camera can compete with the
HST in terms of photometric depth and resolving power in moderately crowded
stellar fields, with smaller investment of telescope time; {\em ii)} Tests
performed with the ROMAFOT package suggest that the data reduction of images
from adaptive optics instrument requires the development of novel techniques to
model the complicated PSF of these imagers.

The MSK is an important feature in the CMD of GCs that can help revising the
age of these fundamental stellar systems. The obvious advantage of using the
$\Delta M({\rm MSTO-MSK})$ approach, being a differential measurement, is that it
is not affected by the errors either in the distance or in the reddening.
This is  reflected in  the significantly smaller error bars in the age
determinations. The analysis presented in this paper reveals that different
photometric bands provide different sensitivity to the method. In this sense,
the sensitivity  decreases for increasing wavelength from the optical to the
near infrared. 

Our data analysis allowed us, using the PISCES data, to measure the MSK along
the  main sequence of M\,15. We use two diagnostics to estimate the absolute age
of this cluster: the magnitude of the MSTO and the magnitude difference $\Delta
M({\rm MSTO-MSK})$. The two methods provide consistent results, and a mean absolute
age of 12.9$\pm$2.6 Gyr and 13.0$\pm$1.1 Gyr, respectively. 

A systematic theoretical analysis of the dependence of the MSK magnitude over a
wide range of ages and metallicities is mandatory to firmly establish the
uncertainties affecting the $\Delta M({\rm MSTO-MSK})$ method. Nonetheless, our
results suggest that using high quality, optical data-bases such as those based
on existing HST data can provide a fundamental starting point to globally revise
the age of the GC systems. The PISCES data presented here soundly demonstrate
the potential of ground-based NIR data using adaptive optics technology to
obtain deep photometry in crowded stellar fields. Moreover, current ongoing
observing facilities at the 10m class telescopes  using either SCAO (FLAO at
LBT) or MCAO (GEMS at Gemini South), are providing  excellent data reaching the
MSK in many Galactic GCs (NGC\,1815,  \citealt{turri14}, Turri et al. 2015). The
same outcome applies to the near future facilities like ERIS at ESO VLT.
Nevertheless, a giant leap forward is foreseeable when the next generation of
extremely large telescope, equipped with NIR detectors and AO systems
\citep[e.g.][]{deep11,greggio12,schreiber14}, will be available: the GMT-Giant
Magellan Telescope\footnote{www.gmto.org/}, the TMT-Thirty Meter
Telescope\footnote{www.tmt.org/}, and the E-ELT- European Extremely Large
Telescope\footnote{http://www.eso.org/sci/facilities/eelt/index.html}).

Realistic predictions according to up-to-date instrumental specifications  for
the E--ELT+Maory+Micado configuration suggest that the predicted limiting
magnitude ($K\sim$27.2 assuming a crowding level expected  for the core of a GC,
e.g. \citealt{deep11}), is significantly fainter than the expected MSK magnitude
for any stellar system in the Milky Way Halo (within 100 Kpc, MSK$\sim$24.5).
This is true also when restricting to a relatively short integration time of
600 s, comparable with that of the PISCES data presented in this paper.
Preliminary analysis suggests the error bar on the age will be smaller than 1
Gyr for the entire sample of GCs. This means that we will be able to
successfully apply this method, for the first time, also to the nearby Local
Group galaxies and their GCs.

\begin{acknowledgements} 

The authors warmly thank Don vandenBerg for sharing the new set of isochrones
prior to publication. Support for this work has been provided by the Education
and Science Ministry of Spain (grant AYA2010-16717). GF has been supported by
the Futuro in Ricerca 2013 (grant RBFR13J716). This work was partially
supported by PRIN--INAF 2011 "Tracing the formation and evolution of the
Galactic halo with VST" (P.I.: M.  Marconi) and by PRIN--MIUR (2010LY5N2T)
"Chemical and dynamical evolution of the Milky Way and Local Group galaxies"
(P.I.: F. Matteucci).

\end{acknowledgements}

\end{document}